\documentclass[a4paper,11pt]{article}
\pdfoutput=1 
\usepackage{slashbox}
\usepackage{jcappub} 

\usepackage{graphicx,amssymb,amsmath,xcolor,float}
\usepackage[normalem]{ulem}
\graphicspath{{figs/}}

\definecolor{mygreen}{rgb}{0.0, 0.45, 0.18}

\newcommand{\okinline}[1]{\textcolor{black}{#1}}

\newcommand{\grinline}[1]{\textcolor{black}{#1}}

\newcommand{\blue}{\color{black}}

\newcommand{\black}{\color{black}}
\usepackage[T1]{fontenc} 

\title{Using Deep Learning to Enhance Event Geometry Reconstruction for the Telescope Array Surface Detector}

\author[a]{D.~Ivanov,}
\author[b,c,d]{O.E.~Kalashev,}
\author[b,e]{M.Yu.~Kuznetsov,}
\author[b]{G.I.~Rubtsov,}
\author[f]{T.~Sako,}
\author[g,h]{Y.~Tsunesada,}
\author[b,f]{Y.V.~Zhezher}

\affiliation[a]{High Energy Astrophysics Institute and Department of Physics and Astronomy, University of Utah, Salt Lake City, Utah, 84112-0830, USA}
\affiliation[b]{Institute for Nuclear Research of the Russian Academy of
	Sciences, Moscow, 117312, Russia}
\affiliation[c]{Moscow Institute for Physics and Technology, 9 Institutskiy per., Dolgoprudny, Moscow Region, 141701 Russia}
\affiliation[d]{Novosibirsk State University, Pirogova 2, Novosibirsk, 630090 Russia}
\affiliation[e]{Service de Physique Th\'eorique, Universit\'e Libre de Bruxelles, Boulevard du Triomphe, CP225, 1050 Brussels, Belgium}
\affiliation[f]{Institute for Cosmic Ray Research, University of Tokyo, Kashiwa, Chiba, 277-8582, Japan}
\affiliation[g]{Graduate School of Science, Osaka City University, Osaka, Osaka, 558-0022, Japan}
\affiliation[h]{Nambu Yoichiro Institute of Theoretical and Experimental Physics, Osaka City University, Osaka, Osaka, 558-8585, Japan}
\emailAdd{kalashev@inr.ac.ru}
\emailAdd{mkuzn@inr.ac.ru}


\abstract{The extremely low flux of ultra-high energy cosmic rays (UHECR) makes their direct observation by orbital experiments	practically impossible. For this reason all current and planned UHECR experiments detect cosmic rays indirectly by observing the extensive air showers (EAS) initiated by cosmic ray particles in the atmosphere. The world largest statistics of the ultra-high energy EAS events is recorded by the networks of surface stations. In this paper we consider a novel approach for reconstruction of the arrival direction of the primary particle based on the deep convolutional neural network. The latter is using raw time-resolved signals of the set of the adjacent trigger stations as an input. The Telescope Array (TA) Surface Detector (SD) is an array of 507 stations, each containing two layers plastic scintillator with an area of $3$ m$^2$. The training of the model is performed with the Monte-Carlo dataset. It is shown that within the Monte-Carlo simulations, the new approach yields better resolution than the traditional reconstruction method based on the fitting of the EAS front. The details of the network architecture and its optimization for this particular task are discussed.
	
}

\makeatletter
\gdef\@fpheader{}
\makeatother
\begin{document}
\begin{flushright}
INR-TH-2020-027
\end{flushright}
	\maketitle
	\flushbottom
	
	\section{Introduction}
	\label{sec:intro}

The ultra-high energy cosmic ray (UHECR) sources identification is one of the most difficult problems of modern astroparticle physics. UHECRs are believed to have an extragalactic origin since the galactic magnetic field is not capable to keep them inside the Milky Way, \blue that was indicated by observations~\cite{Abbasi:2016kgr, Aab:2020xgf}. \black The well-known Hillas criterion~\cite{Hillas:1985is} strictly limits the set of potential UHECR source classes, based on the requirement of a particle's Larmor radius not exceeding the accelerator size $R$, otherwise the particle escapes the accelerator and cannot gain the energy any further. This translates into the condition for maximum acceleration energy $E_{\max}$ as a function of the particle electric charge $q$ and the typical magnetic field strength $B$ in the source environment: $E_{\max} \leq q B R$. Therefore, heavier nuclei can be accelerated to higher energies, however the radiation loss restrictions may also apply~\cite{Ptitsyna:2008zs}. Powerful active galaxies, starburst galaxies and shocks in galaxy clusters are often considered among possible UHECR sources.
 
\begin{figure}[ht]
	\includegraphics[width=30pc]{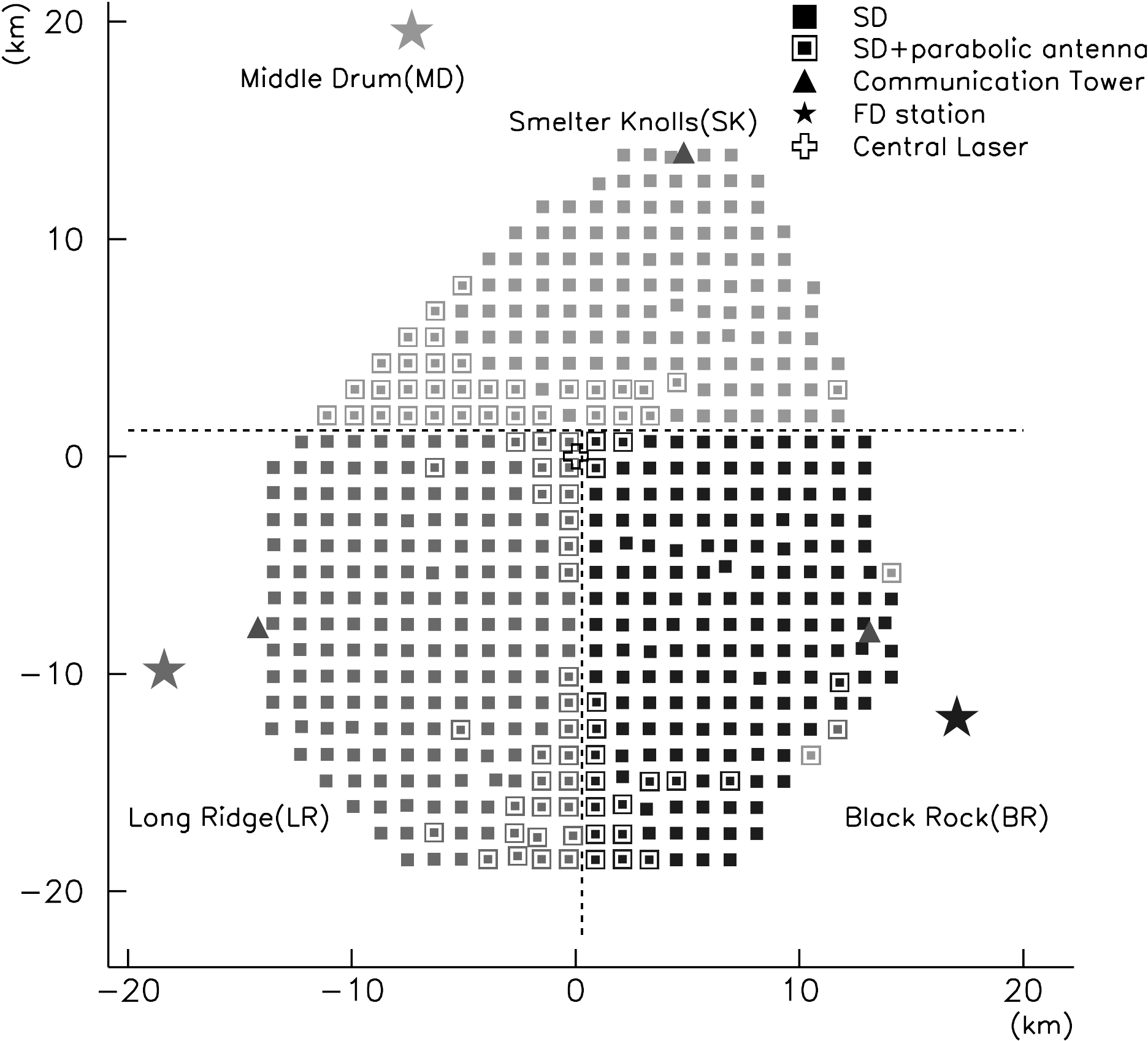}
	\caption{\label{fig:sd_array_map}
	The Telescope Array experiment located in Utah, USA. 507 squares denote surfrace detector stations~\cite{AbuZayyad:2012ru}.}
\end{figure}

\grinline{Extremely low flux of the UHECR does not allow to register them directly. Instead, all present UHECR experiments use indirect detection methods based on the registration of extensive air showers (EAS). The latter are three-dimensional cascades of secondary particles generated by a cosmic-ray particle in the atmosphere.} Two largest UHECR experiments, the Pierre Auger Observatory (Auger)~\cite{ThePierreAuger:2015rma} and the Telescope Array (TA)~\cite{AbuZayyad:2012kk, Tokuno:2012mi} operate in the Southern and Northern Hemisphere respectively. Both experiments combine two shower observation techniques. Firstly, they observe the lateral shower \grinline{particle} density profiles on the ground level using the network of surface detector (SD) stations. Secondly, the shower longitudinal profiles are observed via fluorescent light from the excited nitrogen molecules by the fluorescent detector (FD) telescope stations. Note that since the fluorescent light observation is only possible during the moonless nights  with an additional requirement of the clear weather, the FD duty cycle is approximately $10\%$ of that of the SD. 
TA observatory, the largest experiment in the Northern Hemisphere,  covers the area over 700 $km^2$ in Utah, USA, with over 500 ground scintillation stations, placed at a distance of 1.2 km from each other in rectangular grid and with 3 fluorescence telescope stations placed in the corners  of the SD array\grinline{, see Fig.~\ref{fig:sd_array_map}. The TA SD records an event if three or more adjacent stations are triggered by particles of the EAS. The standard reconstruction procedure estimates the primary particle arrival direction and energy by fitting the arrival times of the shower front and lateral distribution of the signals at the stations, correspondingly.}

\grinline{The shower reconstruction procedure is constructed with use of the Monte-Carlo modelling. The latter requires to simulate particle interactions in the atmosphere which involves} the extrapolation of nucleon-nucleus interaction cross section up to energies of hundreds TeV in center-of-mass system. This introduces unavoidable systematic uncertainty in the estimations of all particle properties. The UHECR arrival direction, estimated using the shower front timing measurements (see below), is the least model-dependent quantity. However, \blue once arrival directions are reconstructed the theoretical interpretation of their distribution in the sky strongly depends on the assumption of the primary particle nature. \black The latter is very difficult to infer from the observable EAS properties, because showers initiated by protons and heavy nuclei are very similar. So far only the average UHECR composition can be estimated based on the large amount of showers observation~\cite{Abbasi:2018nun, Aab:2014kda}. Moreover, the upper limits on the possible admixture of $\gamma$-rays~\cite{Aab:2016agp,Abbasi:2018ywn} and neutrinos~\cite{Aab:2019auo,Abbasi:2019fmh} in the UHECR flux were obtained. Recent studies by Auger~\cite{Aab:2017cgk} indicate that the UHECR mass composition changes from mostly light nuclei around several EeV to the heavier ones above 10 EeV, which means that the highest energy cosmic ray arrival directions may not necessarily point back to their sources due to the substantial (several tens of degrees) cosmic rays deflections in the Galactic magnetic field. This fact makes the source identification task extremely difficult. 

The search of the UHECR sources has been addressed with multiple approaches. One group of these approaches is based on the study of large and intermediate scale anisotropies in the UHECR arrival directions distribution. The results of these methods include indication of the hotspot by TA~\cite{Abbasi:2014lda} and measurement of the dipole by Auger~\cite{Aab:2017tyv, Aab:2018mmi}. Another manifestation of the large scale UHECR anisotropy is a difference between TA and Auger energy spectra~\cite{Abbasi:2018zio}. The latter is not significant in the overlapping region of the sky of two experiments. Though being weakly dependent on UHECR source model and composition the large and intermediate scale anisotropy approach can give only a general picture of what real sources could be~\cite{diMatteo:2017dtg, Wittkowski:2017nfd, AlvesBatista:2017vob}. These results are more demanding for an accuracy of the UHECR energy reconstruction than the accuracy of reconstruction of events arrival directions.

Another group of approaches is focused on the small scale anisotropies and assumes testing some predefined UHECR source model. There exists, however, a freedom in the choice of the source parameters in construction of predefined search catalogue and a blind scanning over these parameters leads to a statistical penalty. 
\black The results of these methods may possess a certain level of ambiguity since there are physically different models of sources that lead to similar UHECR distributions on the sky. Despite these challenges, this method is potentially able to discover the particular UHECR sources. Among the results of these methods application are HiRes evidence of small fraction of UHECR from BL Lacs~\cite{Gorbunov:2004bs, Abbasi:2005qy} as well as a recent Auger result on starburst galaxies~\cite{Aab:2018chp}. Another approach is based on the study of events auto-correlation at small angles and search for spatial doublets or triplets of events
Although the present results on auto-correlation are consistent with the isotropic distribution of events, the lower limits on the density of the UHECR sources are established~\cite{Dubovsky:2000gv,Abreu:2013kif}. The methods of this group can significantly benefit from the improvement of the experiments angular resolution, especially in the searches in which small deflection of the particles from their sources is assumed.

There are also ``hybrid'' approaches to anisotropy search that simultaneously include information about UHECR arrival directions and their energy. Possible applications of these methods include searches for specific energy-angle correlation patterns in the UHECR distribution~\cite{Aab:2020mfn, Abbasi:2020fxl, Kalashev:2019skq, Bister:2020rfv}. The improvement in experiment's angular resolution may boost the analyses of this kind, while the simultaneous improvement in energy resolution may also be required.  

\grinline{The standard reconstruction of the extensive air shower events is well established. It makes use of empirical functions to fit the readings of the SD stations. While each of the TA SD station is recording full time-resolved signal at each of the two scintillator layers, the standard reconstruction is limited with using only two numerical values from each station: first particle arrival time and integral signal. One may expect an improvement of the reconstruction precision if the complete waveform data are used as an input of the procedure. The complexity of the SD signals makes it impractical to process full data set with empirical functions. On the other hand, machine learning methods are shown applicable to large and complex data in astroparticle physics~\cite{Erdmann:2020kfi}. The use of deep convolutional networks for UHECR reconstruction was first demonstrated in Ref.~\cite{Erdmann:2017str} for the experimental conditions of the Pierre Auger observatory.}

\grinline{In this paper we develop new event reconstruction method based on the machine learning algorithms. We illustrate its capacity to achieve higher precision of arrival direction reconstruction compared to the standard reconstruction.} The paper is organised as follows: in Section~\ref{sec:method} we briefly review the standard reconstruction procedure of UHECR properties used in TA SD and then describe a new method to enhance this reconstruction by means of the convolutional neural network (CNN). In Section~\ref{sec:results} we compare the results of the CNN-enhanced reconstruction of UHECR arrival directions with the respective results of the standard TA SD reconstruction. We summarize in Section~\ref{sec:concl}.

\section{Method}
\label{sec:method}

In this work we focus on the event reconstruction of the surface detector array. 
Modern experiments record the full time-resolved signal of each SD station (in case of the Telescope Array in each of the two layers of the scintillator). The traditional analysis methods are based mostly on the values that could be measured by the previous generation of the detectors: the arrival time of the first particle and the integral signal of each detector. The chemical composition analysis performed by the Auger and TA collaborations uses a number of empirically established integral characteristics of the measured signal. In both cases, not all of the data available for the analysis were used. 

The machine learning methods, in particular, deep convolutional neural networks have been very successful in image recognition and many related tasks, including challenges in astroparticle and particle physics. Due to these advancements it is now possible to perform the analysis utilizing all the experimental data available. The method we propose uses existing standard SD event reconstruction as a first approximation. We describe it below.

\subsection{Standard Reconstruction}
\label{sec:reconstr}

\grinline{The standard SD event reconstruction~\cite{AbuZayyad:2012ru} is built on fitting of the individual station readings with the predefined empirical functions. The event geometry} is reconstructed using the arrival times of the shower front measured
by the triggered stations. The shower front is approximated by the empirical functions first proposed by J. Linsley and L. Scarsi~\cite{PhysRev.128.2384}, then modified by the AGASA experiment~\cite{Teshima:1986rq} and fine-tuned to fit the TA data in a self-consistent manner. The integral signals of the individual stations are used to estimate the \grinline{particle density at distance of 800 meters from the core  $S_{800}$ which plays a role of the} lateral distribution profile normalization~\cite{Takeda:2002at}.

The fit of shower front and lateral distribution function is performed with 7 free parameters~\cite{Abu-Zayyad:2013dii}: the position of the shower core ($x_{\mbox{core}}$, $y_{\mbox{core}}$), the arrival direction of the shower in the horizontal coordinates ($\theta$, $\phi$), the signal amplitude at the distance of 800 meters from the shower core $S_{800}$, the time offset $t_0$ and Linsley front curvature parameter $a$~\cite{PhysRev.128.2384}. The following functions \grinline{$t(\vec R)$} and \grinline{$S(r)$} are employed for the shower front and the lateral distribution correspondingly:
\begin{align*}
t \left(\vec R \right) &= t_0 + t_{\mbox{plane}}\left(\vec R \right) + a \times \left(1+r/R_L\right)^{1.5} LDF\left(r \right)^{-0.5},\\
S \left(r \right) &= S_{800} \times LDF\left(r \right),
\end{align*}
\noindent \grinline{where $LDF(r)$ is the empirical lateral distribution profile and $t_{\mbox{plane}}$ is the arrival time of the shower plane at the station with the location given by $\vec R$ in the pre-defined coordinate system of the array centered at the Central Laser Facility (CLF)~\cite{Takahashi:2011zzd}}:
\begin{align*}
LDF\left(r \right) &= \left(\frac{r}{R_m}\right)^{-1.2}\left(1+\frac{r}{R_m}\right)^{-(\eta-1.2)}\left(1+\frac{r^2}{R_1^2}\right)^{-0.6},\\
t_{\mbox{plane}}\left(\vec R \right) &= \frac{1}{c} \vec n \left(\vec R - \vec R_{\mbox{core}}\right),
\end{align*}
\noindent \grinline{where $\vec R_{\mbox{core}}$ is the location of the shower core, $\vec n$ -- unit vector towards the direction of the arrival of a primary particle, $c$ is the speed of light and r is a distance from the station to the shower core:}
\begin{align*}
r &= \sqrt{ (\vec R - \vec R_{core})^2 - (\vec n (\vec R - \vec R_{core}))^2},\\
R_m &= 90.0~\mbox{m},~R_1 = 1000~\mbox{m},~R_L = 30~\mbox{m},\\
\eta &= 3.97 - 1.79 \left(\sec\left(\theta\right) -1\right).
\end{align*}

After the fit is performed, the primary particle energy is estimated as a function
\begin{equation*}
E = E_{SD}(S_{800}, \theta) 
\end{equation*}
of the density $S_{800}$ at the distance of $800$~m from the shower core and the zenith angle $\theta$ using the lookup table obtained with the Monte Carlo simulation~\cite{AbuZayyad:2012ru}. 

In this work we use the full Monte Carlo simulation of TA SD events induced by the protons and nuclei for both the standard and the machine learning-based reconstructions. Primary energies are distributed assuming the spectrum of the HiRes experiment~\cite{Abbasi:2007sv} \blue in $10^{17.5}-10^{20.5}$~eV energy range. \black The events are generated by CORSIKA \grinline{version 7.3500}~\cite{Heck:1998vt} with the EGS4~\cite{Nelson:1985ec} model for the electromagnetic interactions, QGSJET~II-03~\cite{Ostapchenko:2004ss} and FLUKA \grinline{version 2011.2b}~\cite{Fasso:2003xz} for high- and low-energy hadronic interactions. The thinning and dethinnig procedures with parameters described in~\citep{Stokes:2011wf} are used to reduce the calculation time. Both proton and nuclei event sets are sampled assuming isotropic primary flux with zenith angles $\theta < 60^\circ$.
\blue The total of 12642 CORSIKA events were simulated for primary protons, 8549 for helium, 8550 for nitrogen and 8715 for iron.
The showers from the proton and the nuclei libraries are processed by the code simulating the real time calibration detector response by means of GEANT4 package~\citep{Agostinelli:2002hh}. Each CORSIKA event is thrown to the random locations and azimuth angles within the SD area multiple times. The Monte-Carlo simulations are described in details in Refs.~\citep{AbuZayyad:2012ru, 2014arXiv1403.0644T, Matthews:2017waf}. \black
We apply the standard anisotropy quality cuts~\cite{Ivanov:2019icrcSpec} to the simulated events in the same way they are applied to the data. The standard cuts include the constraint on the reconstructed zenith angle $\theta_{rec} < 55^\circ$. \okinline{The total number of the simulated events which pass the quality cuts is about 1.3 million}.

\subsection{Deep Learning-based Event Reconstruction}
\label{sec:ann}
\begin{figure}[ht]
	\includegraphics[width=30pc]{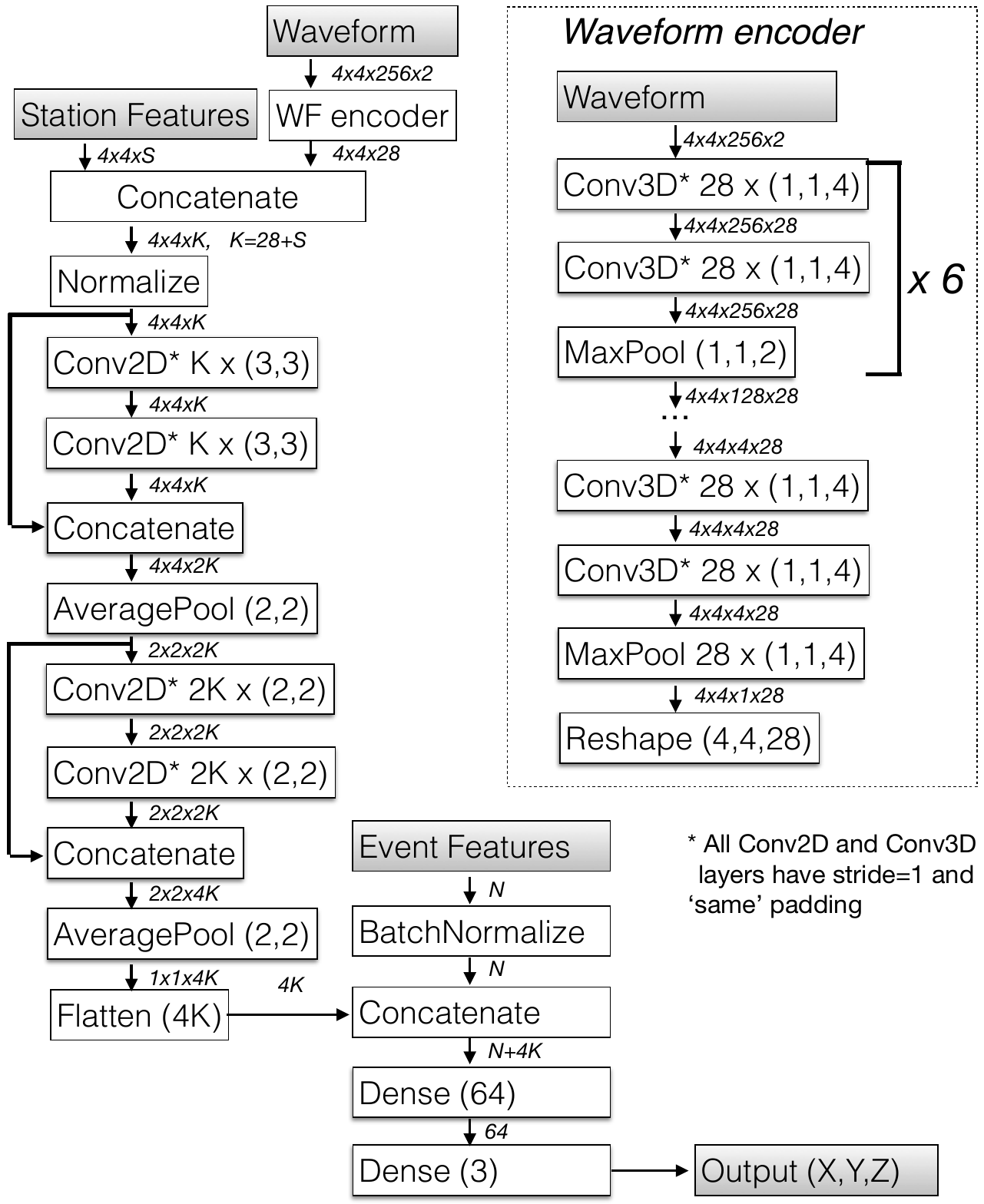}
	\caption{\label{fig:nn_overview} Neural network architecture}
\end{figure}
In essence the method is using the machine learning technique to construct the inverse function for the full detector Monte Carlo simulation. The latter allows to calculate the raw observables as a function of primary particle properties. In case of the Telescope Array surface detector the raw observables are time-resolved signals for the set of the adjacent triggered detectors. The application of this approach to the ultra-high-energy cosmic-ray experiment was first demonstrated in Ref.~\cite{Erdmann:2017str} on toy Monte Carlo simulation roughly following the geometry of Auger observatory. In this work we use the full detector Monte Carlo simulation~\cite{AbuZayyad:2012ru} of the TA observatory to produce the time-resolved signals from the two layers of SD stations. The inverse function is constructed by means of a multi-layer feed-forward artificial neural network (NN). This class of parameterized algorithms is known to be able to approximate any continuous function with any finite accuracy~\cite{HORNIK1991251}. The NN free parameters, weights, are tuned in the so-called training procedure which is in essence the error minimization on a set of training examples. 

In practice, the achievable accuracy is often limited by the stochastic nature of the problem introducing unavoidable bias. In case of EAS development the main source of the stochastic uncertainty is perhaps the first particle interaction which occurs at a random depth. In this work we try to enhance the existing event reconstruction procedure. To rate the relative performance of the enhanced reconstruction with respect to the baseline one the explained variance score could be used
\begin{equation}
EV(y, \hat{y}) = 1 - \frac{Var\{ y - \hat{y}\}}{Var\{y\}}  \label{eq:EV}
\end{equation}
Here $y$ is the true value of the quantity being predicted, which can also be interpreted as the error of the baseline model (zero approximation), and $\hat{y}$ is our estimate of $y$, i.e. the correction to zero approximation calculated in the enhanced model. \okinline{The variance in~\ref{eq:EV} is calculated over the entire test data set.} The ideal reconstuction would have $EV=1$, while the presence of unavoidable bias means that $EV$ is limited from above by some value $EV_{max}<1$. \okinline{If an enhanced reconstruction performs better than the baseline one, then $EV > 0$}. The systematic uncertainty, e.g. hadronic interaction model at ultra-high energies, does not constrain the NN accuracy nominally, however it should be taken into account when interpreting the NN model output. Other than that, the accuracy of the NN model is limited just by the complexity of the network, i.e. the number of trainable weights, with more complex networks requiring more data to train.

For our purpose we utilize the convolutional NN~\cite{CNN}, a special kind of feed-forward neural network which is known to perform well in image or sequence processing. \okinline{Their superior performance compared to other feed-forward networks, such as multi-layer perceptron (MLP) is explained by their ability to efficiently extract and combine local image features on different scales using less free parameters (trained weights). This is achieved by limiting the amount of the surrounding cells to which the next layer neurons are connected and reusing the same weights for different parts of the image. A convolution layer is described by the kernel of fixed size which is less or equal to the size of image. If the size of kernel is equal to the size of image the convolution layer is equivalent to the fully connected perceptron layer. In this work have tried different kernel sizes including the full size of image (or sequence for waveform features extraction).}

\begin{figure}[ht]
	\includegraphics[width=30pc]{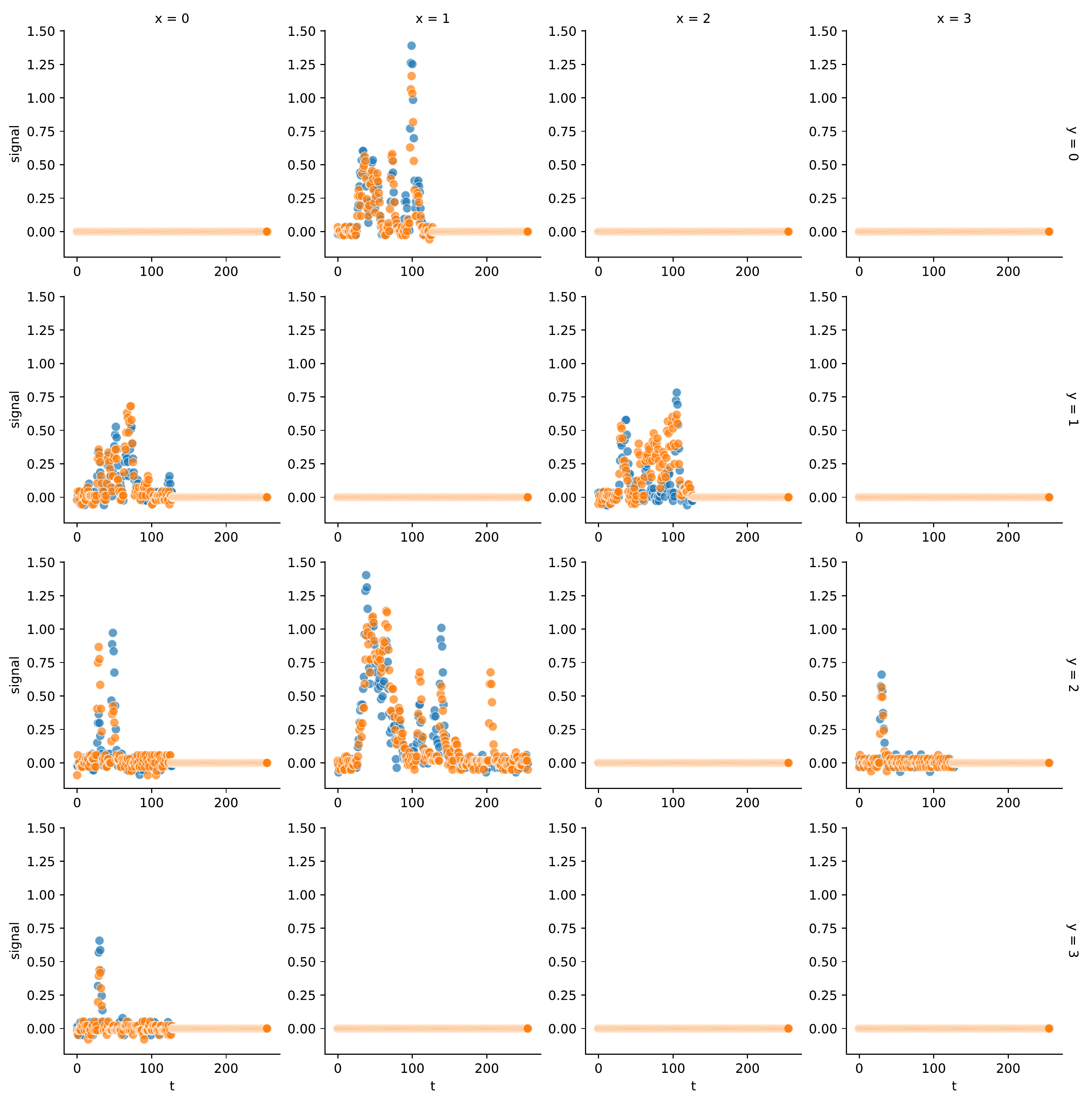}
	\caption{\label{fig:ml_event_example}  \grinline{An example surface detector event in a form used as an input for NN reconstruction. The waveforms of 4x4 neighboring stations are shown with blue circles for lower layer and orange circles for upper layer. The time unit is 20\,ns.}}
\end{figure}
\okinline{In Fig.~\ref{fig:nn_overview} we show the optimal NN architecture which we arrived to.} The typical UHECR event triggers from 5 to 10 neighbour stations. The readouts from $4\times4$ or $6\times6$ stations around the event core are used as an input with each pixel corresponding to a particular station. There are two time-resolved signals, one per layer, for each \grinline{station}. The typical length of the signal recorded with $20$~ns time resolution does not exceed $256$ points. An example input event is shown in Fig.~\ref{fig:ml_event_example}. The overall dimensionality of the raw input data is therefore $4\times4\times256\times2$ or $6\times6\times256\times2$. The position of the \grinline{event core} is estimated using the standard reconstruction. 

The full time-resolved signal from the two scintillator layers of each surface detector \grinline{station} is first converted to a $28$-dimensional feature vector using waveform encoder, consisting of six convolutional layers followed by the max-pooling. \okinline{The number of the waveform features was one of the parameters being optimized with the range from 8 to 64. The waveform encoder kernel size was varied in the range 3-8. In addition, for the waveform feature extraction we tried to use single or double layer perceptron which resulted in worse performance. The scaled exponential linear units are used as the activations to achieve the self-normalizing property~\cite{selu_alpha_dropout}. Slightly \grinline{inferior} accuracy was achieved with the \emph{ReLU} activation function.}

The extracted features from the waveforms are treated as a multichannel image using a sequence of 2D convolution layers. We also add several extra \grinline{station (or station signal)} properties $S$ to the set of the extracted signal features before feeding them into the 2D convolution model.  \grinline{These properties include 3 detector coordinates in form of an offset from the perfect rectangular grid, the detector on/off state and saturation flags, integral signal and signal onset time relative to the plane shower front.}

\okinline{In order to compensate the missing or switched off detectors, we have introduced a special \emph{Normalize} layer, somewhat similar to the \emph{Dropout}~\cite{dropout}. Namely, it drops the pixels corresponding to the missing detectors and multiplies the activations of the present detectors by a factor of $N_{total}/(N_{total}-N_{missing})$. Unlike \emph{Dropout} the above layer works exactly the same way in training and inference mode. We found that this trick enhances the explained variance~(\ref{eq:EV}) on the test data by few percent.}

The 2D-convolutional model part consists of two blocks built from the two convolutional layers with $3\times3$ and $2\times2$ kernels followed by max-pooling. \okinline{Kernel sizes were varied from 2 to 4 and the number of convolutions between successive pooling operations was varied from 1 to 3.} The input signal in each convolutional block is concatenated with the output before pooling operation, which facilitates reusing the extracted features at different scales. \okinline{This trick gives another few percent enhancement in terms of the explained variance.} We also tried to replace classic convolution blocks with the separable convolution as it was suggested in~\cite{Erdmann:2017str} and found no positive effect on the accuracy.

The 2D convolutional model outputs a feature vector which is then processed by the two fully connected layers, the latter being the output layer. \okinline{The optimal number of neurons in the first dense layer was chosen from the range between 16 and 128}. Before the feature vector is passed to the first dense layer, it is concatenated with $N$ extra event properties, such as \grinline{the year,} the season, the time of day\footnote{\okinline{Time dependence may appear in the data because of evolving detector calibration}} and (optionally) the standard reconstruction data  (e.g. $S_{800}$). At this step we've found it useful to add a set of 14 composition sensitive synthetic observables calculated in the standard reconstruction~\cite{Abbasi:2018wlq}. \grinline{The event observables include Linsley shower front curvature parameter, area-over-peak,
$S_b$ parameter of the lateral distribution function, the sum of the signals of all stations of the event, total number of peaks in all the waveforms, number of waveform peaks in the station with the largest signal, asymmetry of the signal in the upper and the lower scintillator layers, see the Appendix of ~\cite{Abbasi:2018wlq} for a complete list.}
The output may contain one or several observables. \okinline{The mean square error is used as a loss function. In this work we use three components of the arrival direction unit vector $X$, $Y$ and $Z$ as the output variables. We also tried to use zenith and azimuth angles $\phi$ and $\theta$ for the output and found it less efficient. We suppose that $X$, $Y$ and $Z$ is a better choice to use for the angular distance minimization with the mean square error loss, since in this case for small errors the loss is proportional to the average square of the angular distance between the true and predicted directions, while in case of spherical coordinates there is a degeneracy in $\phi$ for small zenith angles $\theta$.}

The raw waveform data is converted to the log scale before it is passed as an input to the model. The rest of the input and output data are normalized to ensure their zero mean and unit variance. Finally, for the arrival direction reconstruction we've found it useful to predict the correction to the standard reconstruction instead of its absolute value.

\okinline{The optimal model shown in Fig.~\ref{fig:nn_overview} has about 120 thousand adjustable weights, which we optimize using the adaptive learning rate method Adadelta~\cite{adadelta}. 
We split the data into three parts: $80\%:10\%:10\%$ for training, test and validation purposes. The training set consists of roughly 1 million events. Batch gradient decent technique is used with batch size of 1024 samples roughly 1000 batches are passed to training algorithm during single training step (so called epoch). The model is evaluated on the validation data at the end of each epoch. Training stops if the loss function on the validation data is not improving for more than 5 epochs. The optimization is performed for at most 400 epochs, however usually it stops after less than 100 epochs}. 
We tried to apply several regularization techniques, such as $L2$-regularisation, noise admixture and $\alpha$-dropout~\cite{selu_alpha_dropout} and eventually found the early-stop procedure to be sufficient to avoid overfitting. \okinline{This is expected result for the model proposed since number of free model parameters is roughly 10 times less than the number of training samples.}

\okinline{To justify the choice of the model design we have also trained and evaluated the models not using event features and/or waveform data. The model performance in terms of the average, $(EV_{X} + EV_{Y} + EV_{Z})/3$, explained variance is shown in Table~\ref{table:evNN}. We see that even minimalistic model using only detector features, of which the most important are integral signal and timing, still improves reconstruction accuracy, but utilizing the raw waveform data gives the largest impact on the accuracy.}
\begin{table}[!t]
\begin{center}
\begin{tabular}{|l||*{5}{c|}}\hline
	\backslashbox{Event Features}{Waveform}
	&\makebox[3em]{Included}&\makebox[3em]{Excluded}\\\hline\hline
	Included &0.37&0.22\\\hline
	Excluded &0.30&0.11\\\hline
\end{tabular}
\caption{\okinline{Explained variance of NN-based reconstruction compared to standard reconstruction for models using or not using event observables and waveform data.}}
\label{table:evNN}
\end{center}
\end{table}
The NN model is implemented in \emph{Python}  using \emph{Keras}~\cite{keras} library with the \emph{Tensorflow} backend. We also use \emph{hyperopt} package~\cite{hyperopt} to optimize model hyperparameters, such as dimensionality of the waveform encoder output, the shapes of the convolution kernels and the dense layer size. 

\section{Results}
\label{sec:results}

\begin{figure}[h]
\begin{minipage}{18pc}
	\includegraphics[width=18pc]{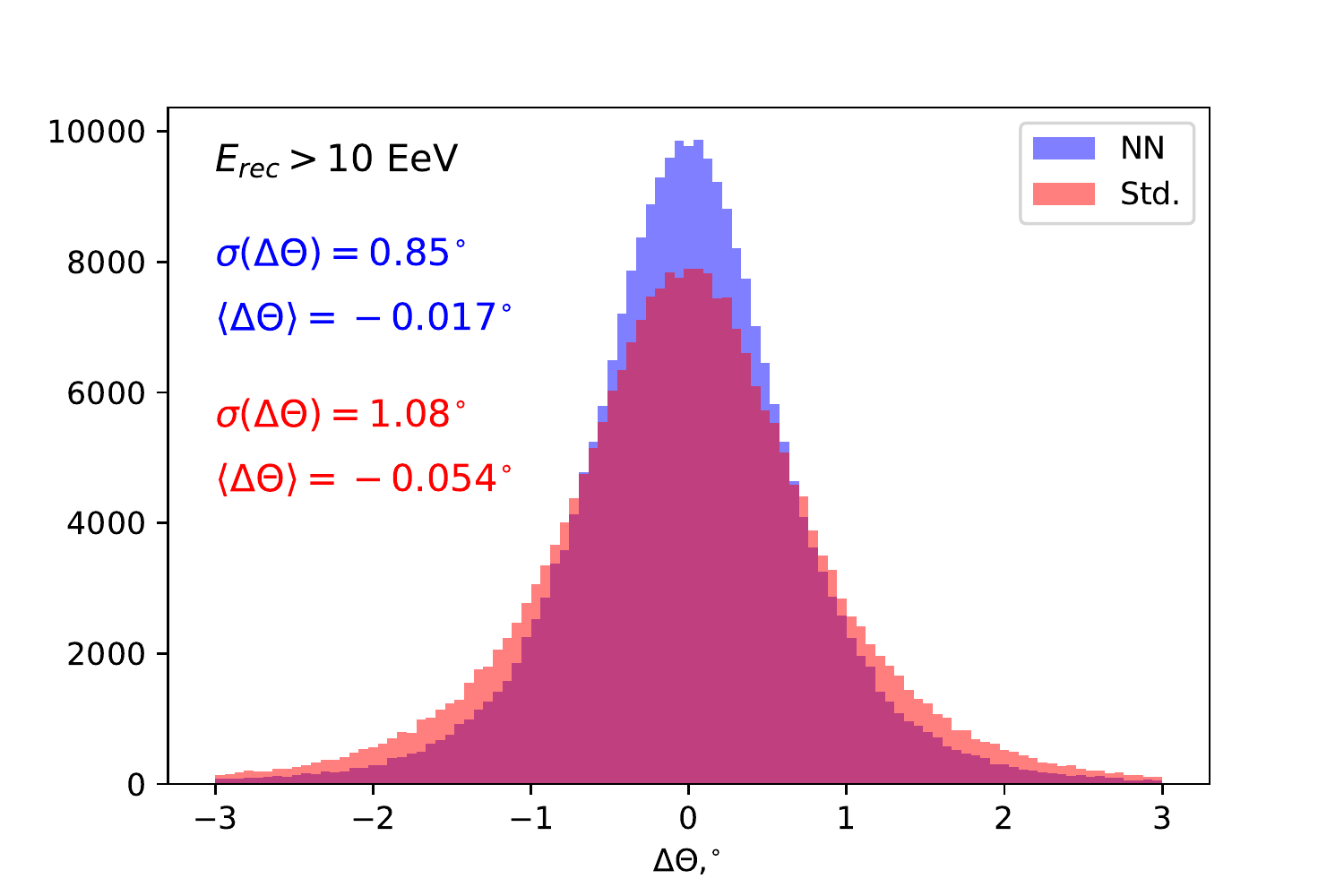}
	\end{minipage}\hspace{2pc}%
	\begin{minipage}{18pc}
		\includegraphics[width=18pc]{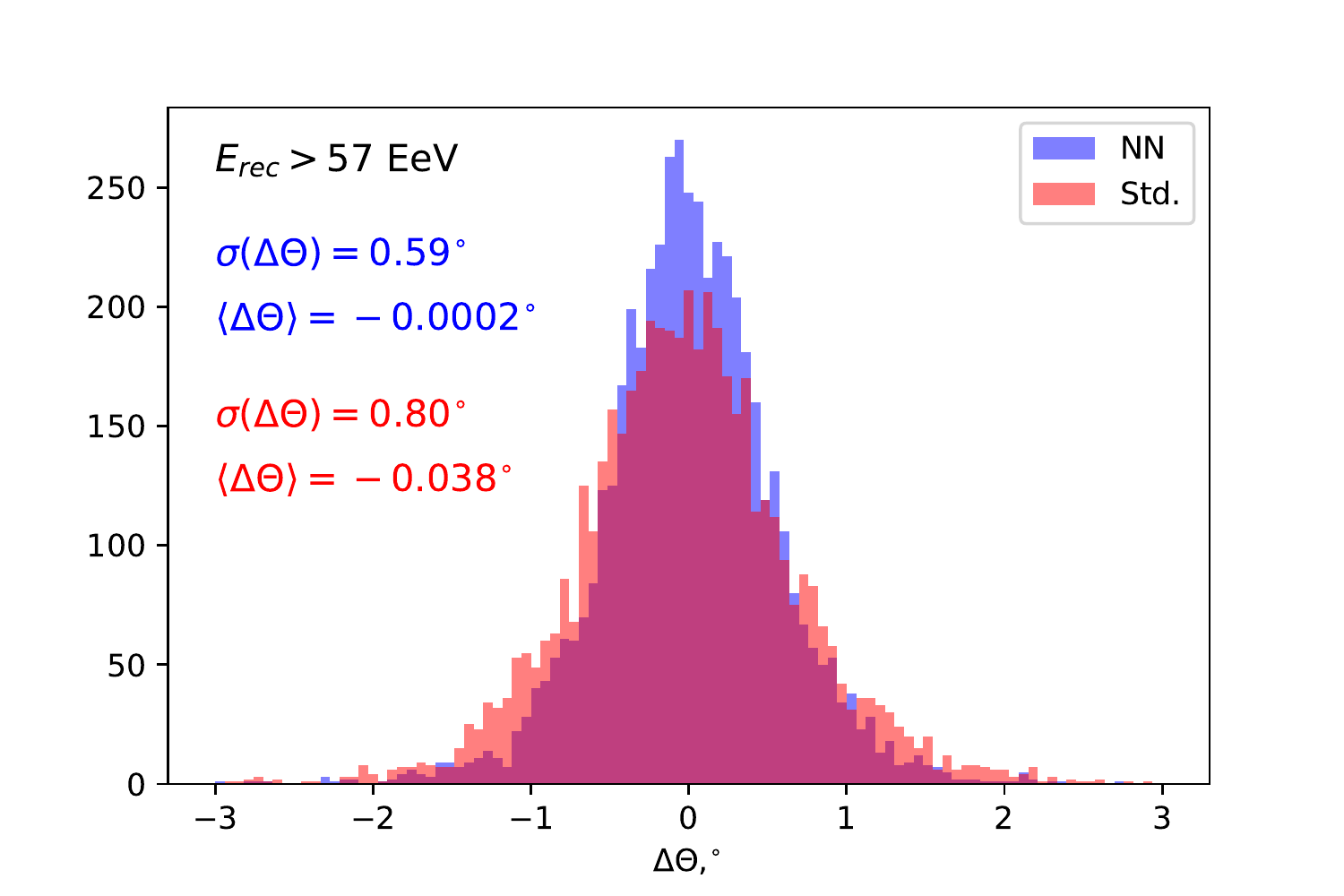}
	\end{minipage}	
\caption{\label{fig:hist_theta} 
Distribution of the difference between the reconstructed and true values of an event's zenith angle for the standard (red histogram) and CNN-enhanced (blue histogram) reconstructions of the proton Monte Carlo event set simulated using QGSJETII-03 hadronic model for the reconstructed energy higher than 10 EeV (left figure) or 57 EeV (right figure).}
\end{figure}

\begin{figure}[h]
\begin{minipage}{18pc}
		\includegraphics[width=18pc]{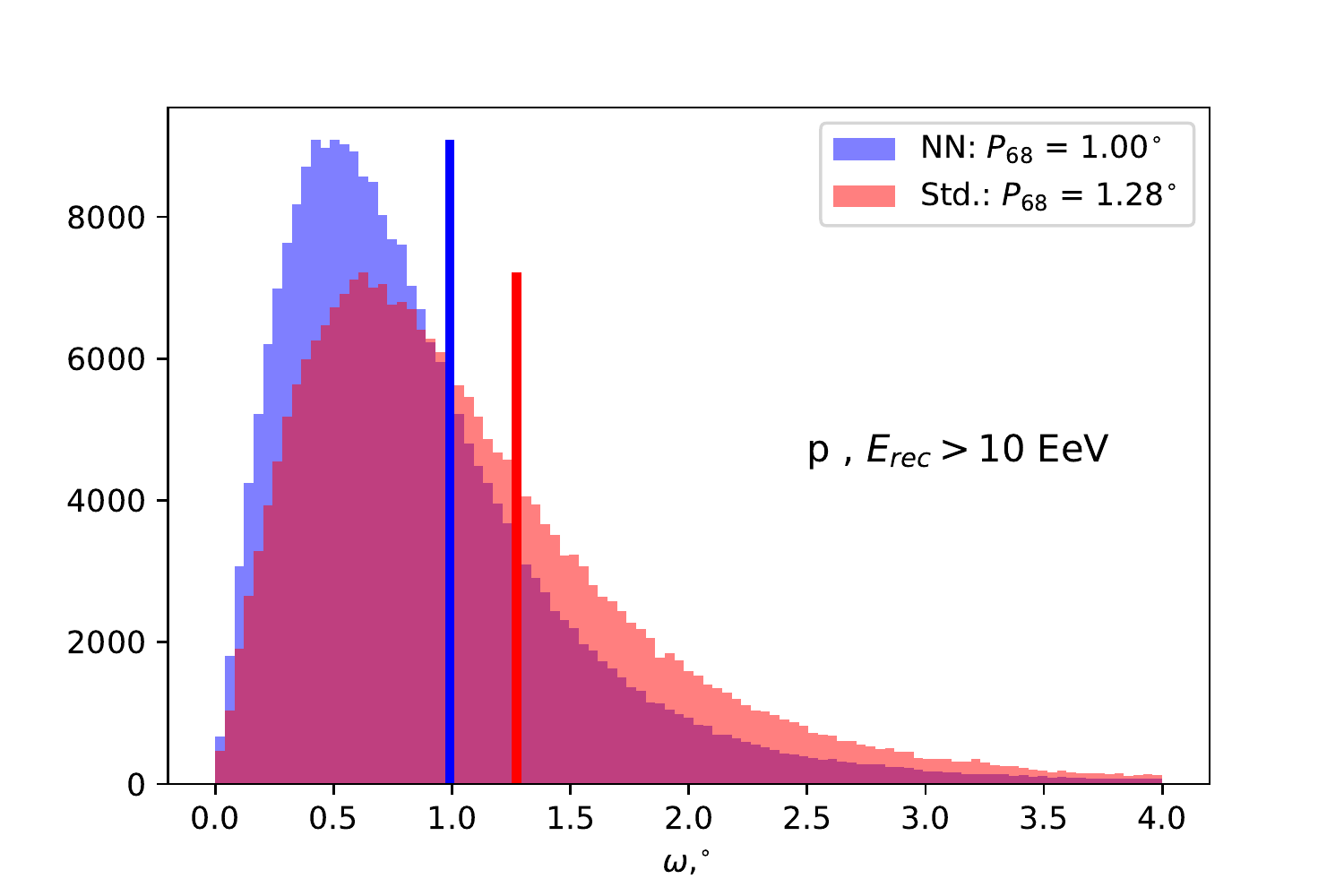}
	\end{minipage}\hspace{2pc}%
	\begin{minipage}{18pc}
	\includegraphics[width=18pc]{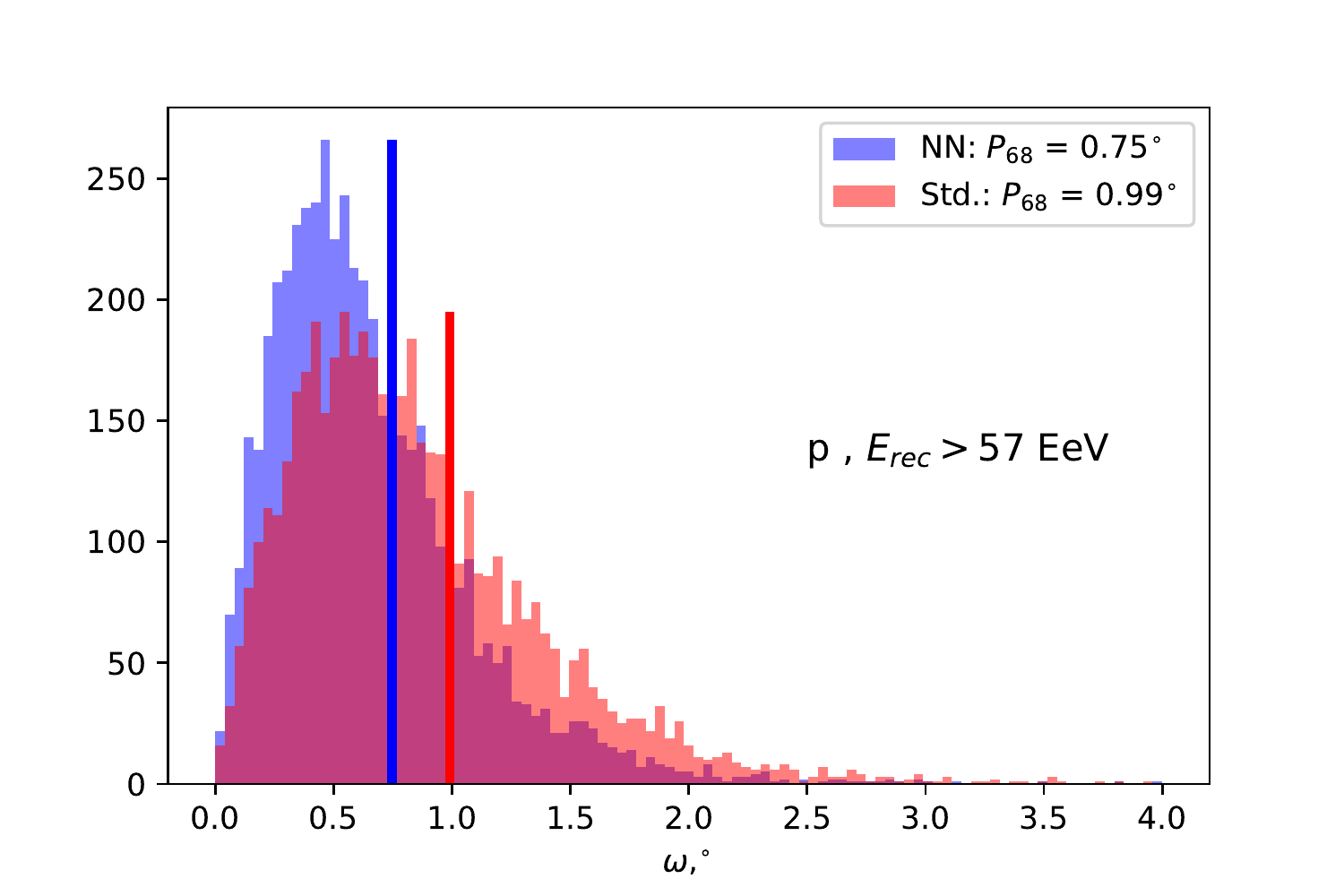}
    \end{minipage} 
\caption{\label{fig:hist_ang_res_nuclei} Angular distance $\omega$ distribution between the true and reconstructed arrival directions for the standard (red histogram) and CNN-enhanced (blue histogram) reconstructions of the proton Monte Carlo event set simulated using QGSJETII-03 hadronic model for the reconstructed energy higher than 10 EeV (left figure) or 57 EeV (right figure). Vertical lines denote the positions of 68\% percentile of the distributions, i.e. the angular resolution values.}
\end{figure}

As the first application of the event's geometry reconstruction we evaluate the event's zenith angle $\theta$. In Fig.~\ref{fig:hist_theta} the accuracy of this observable estimation is compared between the standard and CNN reconstructions. One may see that both the bias and the width of the distribution are smaller for the CNN reconstruction than the standard method. The parameter that is less technical and more interesting for physical applications is the angular resolution of the reconstruction. We calculate the angular resolution as 68\% percentile of angular distance between true and reconstructed cosmic ray arrival direction. The comparison of angular distance distributions of the standard and NN reconstructions for proton Monte-Carlo event set for energies larger than 10 EeV and 57 EeV are shown in Fig.~\ref{fig:hist_ang_res_nuclei} \okinline{along with the angular resolution numbers}.

\begin{figure}[h]
	\begin{minipage}{18pc}
		\includegraphics[width=18pc]{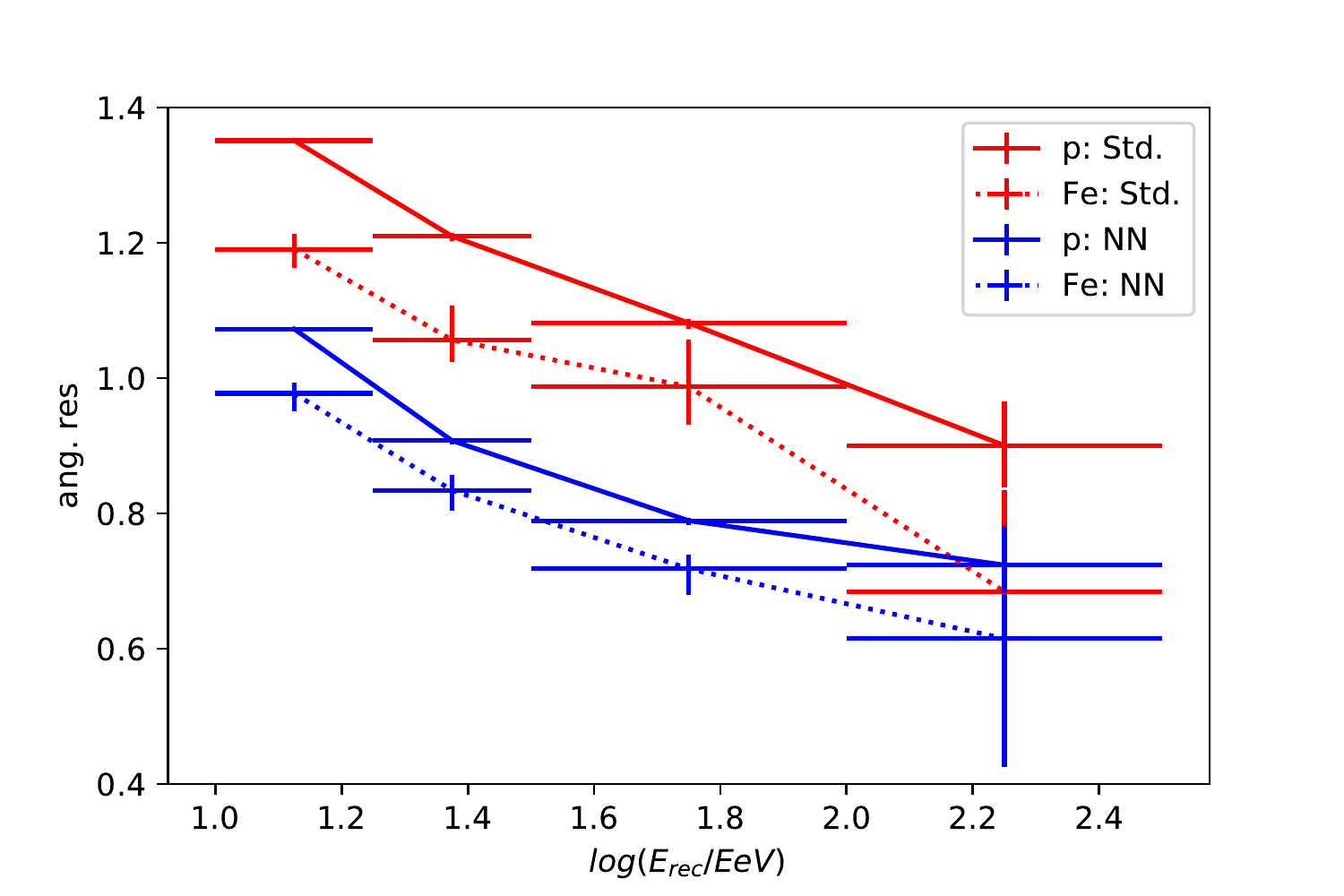}
	\end{minipage}\hspace{2pc}%
	\begin{minipage}{18pc}
		\includegraphics[width=18pc]{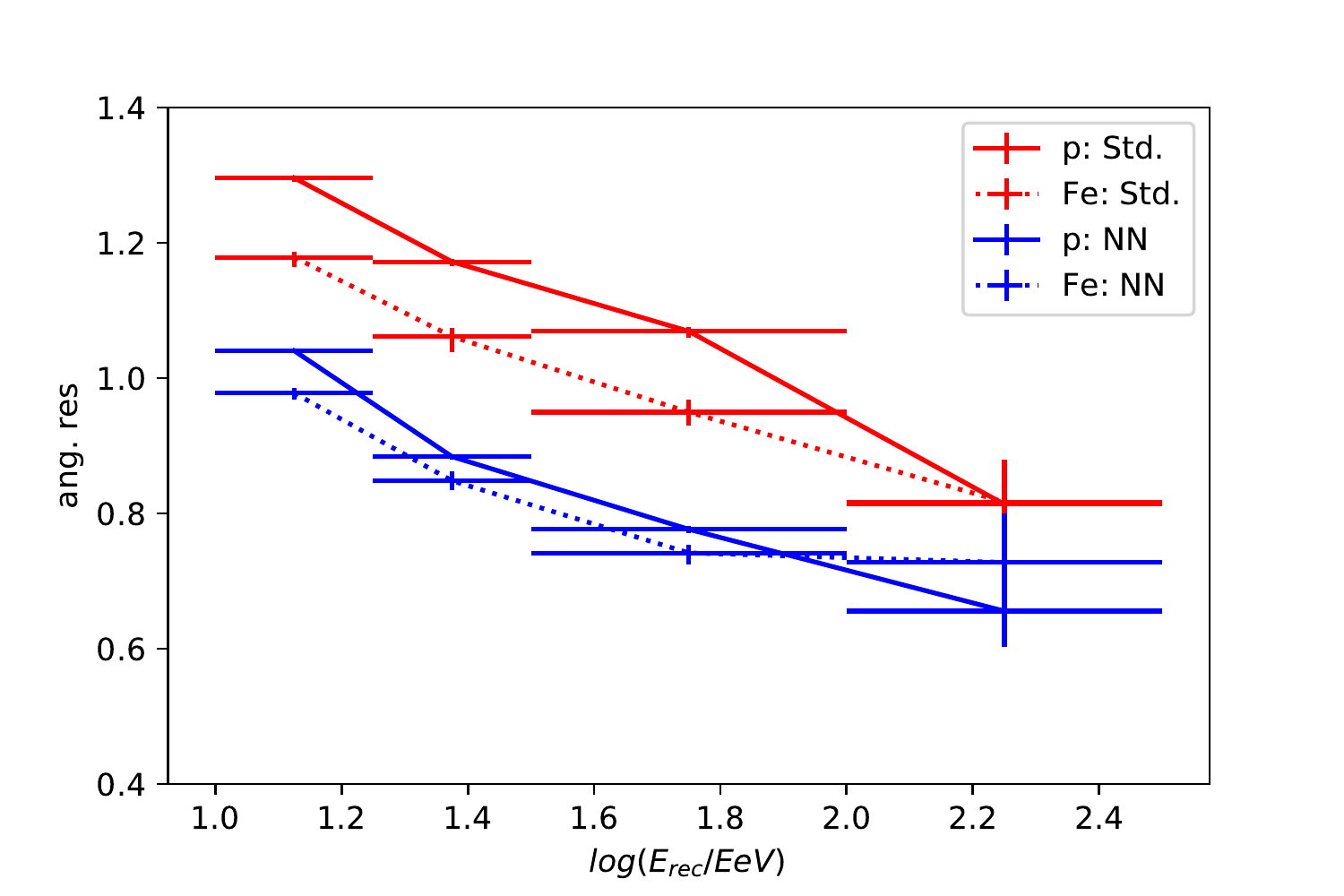}
	\end{minipage} 
\caption{\label{fig:dir_nuclei} Angular resolution for the standard (red curves) and CNN-enhanced (blue curves) reconstructions of the proton (solid lines) and iron (dashed lines) Monte Carlo event sets simulated using QGSJETII-03 (left plot) or QGSJETII-04 (right plots) hadronic interaction models.}
\end{figure}

\okinline{To estimate systematic uncertainty due to the choice of a particular hadronic interaction model we also apply CNN reconstruction trained using QGSJETII-03 model to the data set generated using QGSJETII-04 model}. We plot angular resolutions for the proton and iron Monte Carlo event sets as a function of the reconstructed energy using either QGSJETII-03 or QGSJETII-04 hadronic interaction model \okinline{for test data} in Fig.~\ref{fig:dir_nuclei}. \okinline{In both cases} the training data was composed of Monte Carlo events initialized by H, He, N and Fe nuclei mixture in equal proportions calculated with QGSJETII-03 hadronic interaction model.

\begin{figure}[h]
\begin{minipage}{18pc}
	\includegraphics[width=18pc]{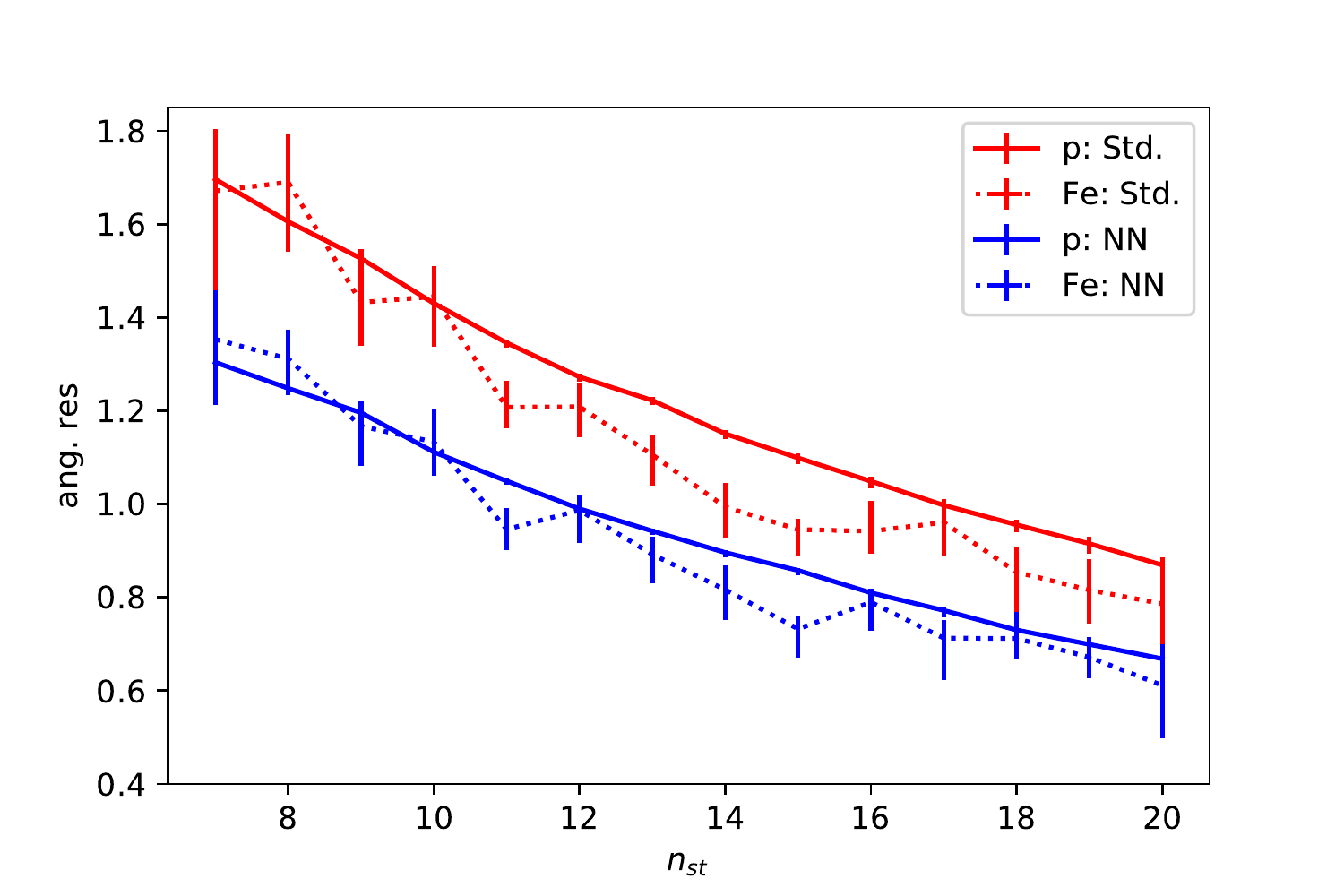}
\end{minipage}\hspace{2pc}%
\begin{minipage}{18pc}
	\includegraphics[width=18pc]{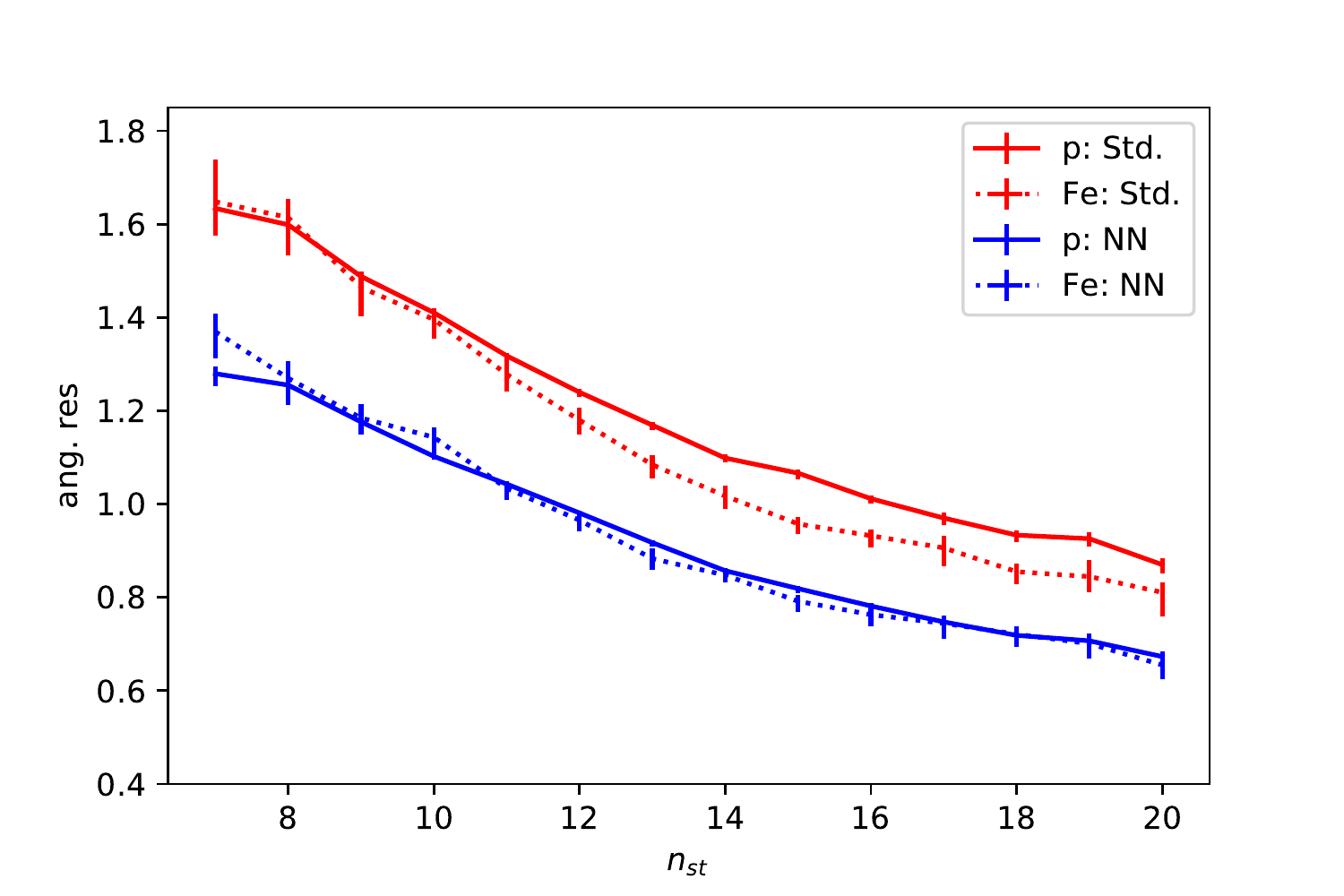}
\end{minipage}
\caption{\label{fig:dir_nuclei_ndet} Angular resolution dependence on the number of detector stations triggered for the proton and iron Monte Carlo sets simulated using QGSJETII-03 (left plot) or QGSJETII-04 (right plot). }
\end{figure}

The angular resolution at a given energy is better on average for heavier nuclei since the EAS produced by heavy nuclei are typically wider and trigger more detectors. This seems to be the main reason of the resolution difference as it is clear from the Fig.~\ref{fig:dir_nuclei_ndet} where the dependence of the resolution on the number of detectors triggered is shown for protons and iron nuclei.

The results presented above were obtained with the NN taking $4\times4$ detector raw signal grid as an input. $6\times6$ architectures show similar performance but take much longer to train. This is expected, since most of the events have just 5-6 neighbor detectors triggered. The larger grid may be useful for the higher energy events reconstruction, where the number of triggered stations is higher.

\section{Conclusion}
\label{sec:concl}

We have shown that the deep learning based methods allow to substantially enhance the accuracy of the TA SD event geometry reconstruction. \okinline{NN architecture and design choices were discussed in detail in Section~\ref{sec:ann}. Table~\ref{table:evNN} confirms the performance of the approach based on the raw waveform usage.}.

The angular resolution for proton induced showers is improved from 1.35$^\circ$ to 1.07$^\circ$ at the primary energy of 1 EeV, \okinline{from 1.28$^\circ$ to 1.00$^\circ$ at the primary energy of 10 EeV and from 0.99$^\circ$ to 0.75$^\circ$ at the primary energy of 57 EeV. The result is especially important for the point source search, since background flux is proportional to the square of the angular resolution.} 

\okinline{Figures~\ref{fig:dir_nuclei} and~\ref{fig:dir_nuclei_ndet} show that} the systematic uncertainties related to the choice of hadronic interaction model which are known to limit the method applicability for the primary particle mass and energy determination seem to be almost irrelevant for the arrival direction reconstruction. We plan to apply the same approach for the photon candidate event reconstruction which is important to improve the directional photon flux limits. We continue the work on the improved method operating with the relaxed quality cuts, which helps to increase the exposure.

\subsection*{Acknowledgments}
We would like to thank Anatoli Fedynitch, John Matthews, Maxim Pshirkov, Hiroyuki Sagawa, Gordon Thomson, Petr Tinyakov, Igor Tkachev and Sergey Troitsky for fruitful discussion and comments. We gratefully acknowledge the Telescope Array collaboration for support of this project on all its stages. The cluster of the Theoretical Division of INR RAS was used for the numerical part of the work. We appreciate the assistance of Yuri Kolesov in configuring the high-performance computing system based on graphic cards. The work is supported by the Russian Science Foundation grant 17-72-20291.
\section*{Data Availability}
\okinline{The data that support the findings of this study are available from the corresponding author upon reasonable request.}

\bibliographystyle{iopart-num}
\bibliography{tasd}

\end{document}